\documentclass[twoside,fleqn]{ActaStyle}
\usepackage{times}

\usepackage{lscape}
\usepackage{graphicx,epsfig}
\usepackage[tbtags]{amsmath}

\def\authorheading{R.~Alkofer et al.}

\def\shorttitle{Infrared Behaviour of Running Coupling}

\begin{document}

\pagerange{1}{8}

{\small {UNITU-THEP-15/2002 \hfill FAU-TP3-02/19}}
\title{ 
The Infrared Behaviour of the Running Coupling in Landau Gauge QCD
\footnote{Talk given by R.A.\ at the conference RENORMALIZATION GROUP 2002,
March 10 - 16, 2002, Strba, Slovakia.}}

\author{
R.~Alkofer\,\email{reinhard.alkofer@uni-tuebingen.de}$^*$, 
C.~S.~Fischer\,\email{chfi@axion01.tphys.physik.uni-tuebingen.de}$^*$,
L.~von~Smekal\,\,\email{smekal@theorie3.physik.uni-erlangen.de}$^\dagger$, 
}
{%
$^*$ Institute for Theoretical Physics, T\"ubingen University \\
Auf der Morgenstelle 14, D-72076 T\"ubingen, Germany\\
$^\dagger$ Institute for Theoretical Physics III, Erlangen University \\
Staudtstr.~7, D-91058 Erlangen, Germany
}

\day{May 7, 2002}

\abstract{%
Approximate solutions for the gluon and ghost propagators as well as
the running coupling in Landau gauge Yang--Mills theories are presented.
These propagators obtained from the corresponding Dyson--Schwinger equations
are in remarkable agreement with those of recent lattice calculations.
The resulting running coupling possesses an infrared fixed point,
$\alpha_S(0) = 8.92/N_c$  for all gauge groups SU($N_c$). Above one GeV the
running coupling rapidly approaches its perturbative form.
}


\section{Motivation: Some aspects of confinement}
\label{sec:intr} \setcounter{section}{1}\setcounter{equation}{0}

The investigation of Quantum Chromo Dynamics (QCD) over the last decades made
obvious that a description of hadrons and their processes from the dynamics of
confined quarks and gluons in one coherent approach is an outstanding
complicated task. 
Knowledge of a completely non-perturbative running coupling would be a very
important step to make towards such a QCD-based description of hadrons.

Given the experimental results in hadron physics it is evident that baryons and
mesons are not elementary particles in the naive sense of the word
``elementary''. The partonic substructure of the nucleon has been determined 
to an enormous precision leaving no doubt that the parton picture emerges from
quarks and gluons, the elementary fields of QCD. This contrasts the well-known 
fact that these quarks and gluons have not been detected outside hadrons. This
puzzle was given a name: {\em confinement}. The confinement hypothesis was
formulated  several decades ago, nevertheless, our understanding of the
confinement mechanism(s) is still not satisfactory. 
Moreover, in contrast to other
non-perturbative phenomena in QCD ({\it e.g.}, dynamical breaking of chiral
symmetry, $U_A(1)$ anomaly, and formation of relativistic bound states), 
it seems not even clear, at present, whether the phenomenon of confinement is
at all compatible with a description of quark and gluon correlations in
terms of local fields in the usual sense of quantum field theory.

One important feature of QCD is asymptotic freedom:  Employing the
renormalisation group (RG) in perturbative calculations clearly leads to an
understanding of the experimental observation that partons interact at large 
(spacelike) momentum transfer only weakly. On the other hand, RG based
arguments prove that  perturbation theory is insufficient to account for
confinement in four-dimensional field theories: Confinement requires the
dynamical generation of a physical mass scale. In presence of such a mass
scale, however, the RG equations imply the existence of essential singularities in physical
quantities, such as the $S$-matrix, as functions of the coupling at $g =
0$. This is because the dependence of the RG invariant confinement scale
on the coupling and the renormalisation scale $\mu$ near the ultraviolet
fixed point is determined by \cite{Gross:1974jv}
\begin{equation}
  \Lambda = \mu \exp \left( - \int ^g \frac {dg'}{\beta (g')} \right)
  \stackrel{g\to 0}{\rightarrow } \mu  \exp \left( - \frac 1 {2\beta_0g^2}
\right),    \quad \beta_0>0 .
  \label{Lambda}
\end{equation}
Therefore a study of the infrared behaviour of QCD amplitudes requires
non-perturbative methods. In addition, as infrared singularities are
anticipated, a formulation in the continuum is desirable. One promising
approach to non-perturbative phenomena in QCD is provided by studies of
truncated systems of its Dyson--Schwinger equations,
the equations of motion for QCD Green's functions, for recent reviews
see {\it e.g.} \cite{Alkofer:2001wg}. As we will see in the following these
studies also allow to extract a non-perturbative running coupling.
One word of warning is, however, in order: 
A non-perturbative running coupling is not a uniquely defined object. 
The extension of the running coupling into the infrared domain requires in the
first step its definition from a specifically chosen Green's function. Of
course, within a chosen gauge the result should be unique, and one should be
able to prove this uniqueness from the Slavnov--Taylor identites of QCD. 
Noting that Green's function are not gauge invariant it is not at all obvious
whether the comparison of non-perturbative running couplings defined in
different gauges can be meaningful at all. 

\section{A possible definition of the non-perturbative running coupling in 
Landau gauge}
\label{sec:Def}

As stated above, the definition of the non-perturbative running coupling 
rests on a specifically chosen Green's function. To this end we note that 
the ghost-gluon vertex in Landau gauge acquires no independent renormalisation,
in specialist language $\widetilde Z_1= 1$. This relates the charge
renormalisation constant $Z_g$ to the ones for the gluon and ghost wave
functions, $1=\widetilde Z_1= Z_g\sqrt{Z_3} \widetilde Z_3$: The gluon leg
provides a factor $\sqrt{Z_3}$, the two ghost legs $\widetilde Z_3$.
As we will demonstrate in the following this allows for a definition of the
running coupling resting solely on the properties of the gluon and ghost
propagators.

In linear covariant gauges the gluon propagator is of the form\footnote{As usual
in these studies a Wick rotation to Euclidean space has been employed.}
\begin{equation}
D_{\mu\nu \; \text{Gluon}} = \frac{{Z}(k^2)}{k^2}
                        \left( \delta_{\mu\nu} - \frac{k_\mu k_\nu}{k^2} \right)
                        + \xi \frac{k_\mu k_\nu}{k^4} .
\end{equation}
The Lorentz condition $\partial_\mu A_\mu^a = 0 $ is strictly implemented only
in the limit $\xi \to 0$ which defines the Landau gauge. Then the gluon
propagator is strictly transverse with respect to the gluon momentum $k$. The
gluon dressing, also in the non-perturbative domain, is solely described by the
function ${Z}(k^2)$. The Faddeev--Popov ghosts are introduced in the process of
quantization: The functional integral over these ghosts is a representation  of
the Jacobian factor induced in the generating functional when enforcing the 
gauge condition. The scalar ghost fields belong to the trivial representation
of the connected part of the Lorentz group. As local fields with space-like
anti-commutativity, they violate the spin-statistics theorem and are thus
necessarily unphysical. The general form of their propagator reads
\begin{equation}
D_{\text{Ghost}} = { { -}} \; \;
    \frac{{G}(k^2)}{k^2}.
\end{equation}
A comparison to the tree level form of these propagators reveals that 
${Z}(k^2) \to const.$ and $G(k^2) \to const.$, up to the perturbative
logarithms,  
for asymptotically large momenta $k^2$.

In the next step we discuss the employed non-perturbative subtraction scheme
and its relation to the definition of the running coupling.
As already stated, the starting point is the following identity for the 
renormalisation constants
\begin{equation}
  \widetilde{Z}_1 \, = \, Z_g Z_3^{1/2} \widetilde{Z}_3 \, = \, 1 \; ,
  \label{wtZ1}
\end{equation}
which holds in Landau gauge.
It follows that the product
$g^2 Z(k^2) G^2(k^2)$ is RG invariant. In absence of any
dimensionful parameter this (dimensionless) product is therefore a function
of the running coupling $\bar g$,
\begin{equation}
  g^2 Z(k^2) G^2(k^2) = f( \bar{g}^2(t_k, g)) \; , \quad
  t_k = \frac{1}{2} \ln k^2/\mu^2 \; .
  \label{gbar}
\end{equation}
Here, the running coupling $\bar g(t,g)$ is the solution of
$d/dt \, \bar g(t,g) = \beta(\bar g) $ with $\bar g(0,g) = g$ and the
Callan--Symanzik $\beta$-function $\beta (g) = - \beta_0 g^3 + {\mathcal
O}(g^5)$. The perturbative momentum subtraction scheme is asymptotically
defined by $f(x) \to x$ for $ x\to 0$. This is realized by independently
setting
\begin{equation}
  Z(\mu^2) = 1 \quad \mbox{and} \quad G(\mu^2) = 1
  \label{persub}
\end{equation}
for some asymptotically large subtraction point $k^2 = \mu^2$.
If the quantity
$g^2 Z(k^2) G^2(k^2)$ is to have a physical
meaning, {\it e.g.}, in terms of a potential between static colour sources, it
should be independent under changes $(g,\mu) \to (g',\mu')$ according to
the RG for arbitrary scales $\mu'$. Therefore,
\begin{equation}
  g^2 Z({\mu'}^2) G^2({\mu'}^2) \, \stackrel{!}{=} \,  {g'}^2 = \bar g^2(\ln
(\mu'/\mu) , g) \; , \label{gbar'}
\end{equation}
and, $f(x) \equiv x $, $ \forall x$. This can thus be adopted as a physically
sensible definition of a non-perturbative running coupling in the Landau
gauge.
In the scheme summarized in the next section, 
it is not possible to realize $f(x) \equiv x$ by simply
extending the perturbative subtraction scheme (\ref{persub}) to arbitrary
values of the scale $\mu$, as this would imply a relation between the functions
$Z$ and $G$ which is inconsistent with the leading infrared behaviour
of the solutions as will become evident from the discussion presented in the
next section. 
For two independent
functions the condition (\ref{persub}) is in general too restrictive to be
used for arbitrary subtraction points. Rather, in extending the perturbative
subtraction scheme, one is allowed to introduce functions of the coupling
such that
\begin{equation}
  Z(\mu^2) \, = \, f_A(g) \quad \hbox{and} \quad G(\mu^2) \, = \, f_G(g)
  \quad \hbox{with} \quad f_G^2 f_A \, = \, 1 \; ,
  \label{npsub}
\end{equation}
and the limits $ f_{A,\, G} \to 1 \, , \; g \to 0 $. Using this it is
straightforward to see that for $k^2 \not= \mu^2$ one has ($t_k = (\ln
k^2/\mu^2)/2$),
\begin{eqnarray}
  Z(k^2) &=& \exp\bigg\{ -2 \int_g^{\bar g(t_k, g)} dl \,
  \frac{\gamma_A(l)}{\beta(l)} \bigg\} \, f_A(\bar g(t_k, g)) \; ,
  \label{RGsol} \\
   G(k^2) &=& \exp\bigg\{ -2 \int_g^{\bar g(t_k, g)} dl \,
  \frac{\gamma_G(l)}{\beta(l)} \bigg\} \, f_G(\bar g(t_k, g)) \; .
  \nonumber
\end{eqnarray}
Here $\gamma_A(g)$ and $\gamma_G(g)$ are the anomalous dimensions
of gluons and ghosts, respectively, and $\beta(g)$ is the Callan--Symanzik
$\beta$-function. Eq. (\ref{wtZ1}) corresponds to the following identity
for these scaling functions in Landau gauge:
\begin{equation}
  2 \gamma_G(g) \, +\, \gamma_A(g)  \, = \, -\frac{1}{g} \, \beta(g)  \; .
  \label{andim}
\end{equation}
One thus verifies that the product $g^2 Z G^2$ indeed gives the running
coupling ({\it i.e.}, Eq. (\ref{gbar}) with $f(x) \equiv x$). Perturbatively,
at one-loop level Eq. (\ref{andim}) is realized separately, {\it i.e.},
$\gamma_G(g) = - \delta \, \beta(g) /g$ and $ \gamma_A(g) = - (1-2\delta)\,
\beta(g)/g$ with $\delta = 9/44$ for $N_f=0$ and arbitrary
$N_c$. Non-perturbatively one can still separate these contributions from the
anomalous dimensions by introducing an unknown function $\epsilon(g)$,
\begin{equation}
  \gamma_G(g) \, =:\, - (\delta + \epsilon (g) ) \,
  \frac{\beta(g)}{g}
  \; \Rightarrow \quad  \gamma_A(g) \, =\, - (1 - 2\delta - 2\epsilon (g)
  ) \, \frac{\beta(g)}{g}  \;  . \label{RGgam}
\end{equation}
This allows to rewrite Eqs.~(\ref{RGsol}) as follows:
\begin{eqnarray}
  Z(k^2) &=& \biggl( \frac{\bar g^2(t_k, g)}{g^2} \biggr)^{1-2\delta}
  \exp\bigg\{ - 4\int_g^{\bar g(t_k, g)} dl \, \frac{\epsilon(l)}{l} \bigg\}
  \, f_A(\bar g(t_k, g)) \; , \label{RGsol1} \\
  G(k^2) &=& \biggl( \frac{\bar g^2(t_k, g)}{g^2}  \biggr)^{\delta}  \,
  \exp\bigg\{ 2 \int_g^{\bar g(t_k, g)} dl \, \frac{\epsilon(l)}{l} \bigg\}
  \, f_G(\bar g(t_k, g)) \; .
  \nonumber
\end{eqnarray}
This is also possible in the presence of quarks. In
this case one has
$\delta = \gamma_0^G/\beta_0 = 9N_c/(44 N_c - 8 N_f)$ for $N_f$ flavours in
Landau gauge. The above representation of the renormalisation functions
expresses clearly that regardless of possible contributions from the unknown
function $\epsilon(g)$, the resulting exponentials cancel in the product $G^2
Z$. For a parameterisation of the renormalisation functions, these
exponentials can of course be absorbed by a redefinition of the functions
$f_{A,\, G}$. The only effect of such a redefinition is that the originally
scale independent functions $f_{A,\, G} (\bar g(t_k, g))$ will acquire a
scale dependence by this, if $\epsilon \not= 0$.

For the truncation scheme presented in the next section
it is possible, however, to obtain
explicitly scale independent equations thus showing that the solutions for the
renormalisation functions $G$ and $Z$ obey one-loop scaling at all scales
\cite{vonSmekal:1997is}. 
In particular, this implies that the products $g^{2\delta} G$
and $g^{2(1-2\delta)}
Z$ are separately RG invariants  (as
they are at one-loop level). As for the renormalisation scale dependence, the
non-perturbative nature of the result is therefore buried entirely in the
result for the running coupling.

\section{Infrared exponents for gluons and ghosts}
\label{sec:IRexp}

Having at hand a non-perturbative definition of the running coupling 
which employs only the properties of the gluon and ghost propagators,
an equally non-perturbative method is then required to determine these
propagators. 
Their infrared singularity structures are particularly well accessible  
by non-perturbative continuum methods. 
To this end the Landau gauge Dyson--Schwinger
equations (DSEs) have been solved analytically in the infrared  
\cite{vonSmekal:1997is,Atkinson:1998tu,Zwanziger:2001kw,Lerche:2001,
Fischer:2002eq,Fischer:2002hn}. Necessarily, however, DSE studies  
will always be subject to truncations in order to obtain
the closed system of equations to be studied. Thus, the quality of such a
study crucially depends on the justification of the truncations.
Some uncertainty about these truncations remains, however, even in the most
ambitious study.    
On the other hand, corresponding lattice calculations, see {\it e.g.}  
\cite{Bonnet:2000kw,Langfeld:2001cz,Langfeld:2002},
include all non-perturbative physics but are limited for
small momenta by the finite lattice volume. 
Despite their respective shortcomings both these
approaches agree encouragingly well in the general observations: 
there is clear
evidence for an infrared finite or even vanishing gluon propagator and a
diverging ghost propagator. This is in accordance with the Kugo--Ojima
confinement criterion, which in Landau gauge includes the statement that the
ghost propagator should  be more singular than a simple pole
\cite{Kugo:1995km}.

\begin{figure}
  \centerline{ \epsfig{file= 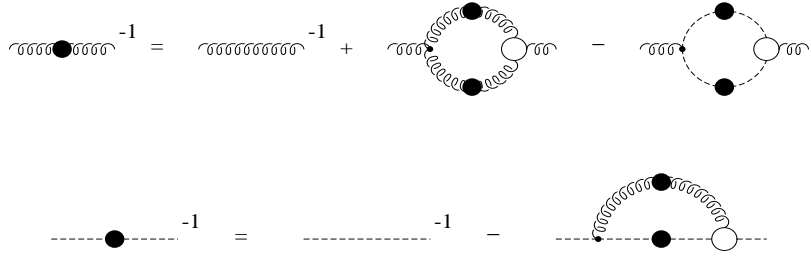,
                                            width=0.8\linewidth,height=35mm} }
  \vskip 3mm
  \caption{Diagrammatic representation of the truncated gluon and ghost
  Dyson--Schwinger equations studied in this letter. Terms with four--gluon vertices
  have been dismissed.}
  \label{GluonGhost}
\end{figure}

For the purpose of this talk we will concentrate on the truncation scheme
for DSEs presented in refs.\ \cite{Fischer:2002eq,Fischer:2002hn}.
The infrared behaviour of the ghost and gluon system 
was studied analytically
for a wide class of non-perturbatively dressed ghost-gluon vertex functions
in ref.\ \cite{Lerche:2001}.  
Based on few and general assumptions about the generic structure of this 
vertex, it was thereby concluded that its dressing should not affect the
qualitative findings. We therefore restrict to bare three-point vertices for
simplicity.  
All contributions from explicit four-gluon vertices are neglected in addition.
A diagrammatical representation of the resulting system of equations
is presented in Fig.~\ref{GluonGhost}.
This truncation scheme provides the correct anomalous dimensions of the ghost
and gluon dressing functions, $Z(k^2)$ and $G(k^2)$, in the ultraviolet region
of momentum. On the other hand, it reproduces the infrared exponents found in
\cite{Zwanziger:2001kw,Lerche:2001} which are close to the ones extacted from
lattice calculations \cite{Bonnet:2000kw,Langfeld:2001cz,Langfeld:2002}.
The numerical solution to the truncated DSEs for the gluon and the ghost 
propagators\footnote{A detailed description of corresponding numerical
techniques can be found in refs.\ \cite{Hauck:1998sm,Atkinson:1998tu,
Fischer:2002eq}.} proved to be compatible with only one out of the two
solutions reported in the infrared analysis of ref.~\cite{Zwanziger:2001kw}. 
This thus demonstrates that not every analytical solution for asymptotically
small momenta necessarily connects to a numerical solution for finite momenta.

\begin{figure}
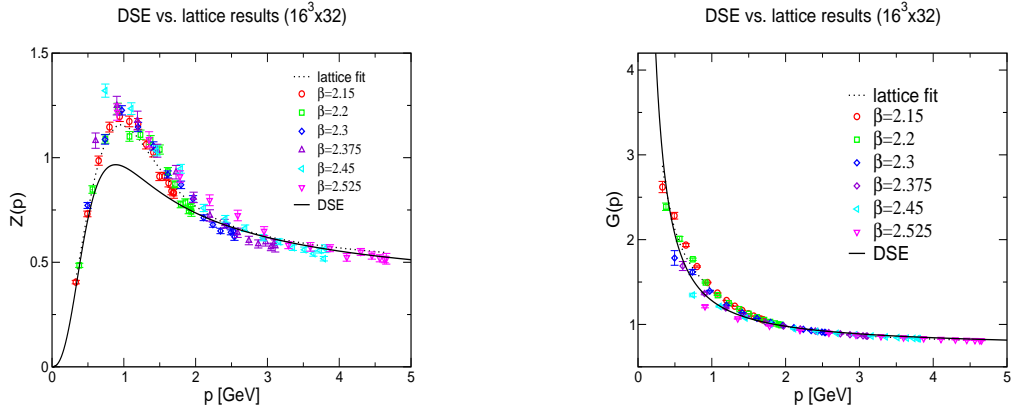

\centerline{
\epsfig{file=gluon.eps,width=0.4\linewidth,height=0.4\linewidth}
\hfill
\epsfig{file=ghost.eps,width=0.4\linewidth,height=0.4\linewidth}
}
\caption{\label{lattice.dat}
Solutions of the Dyson-Schwinger equations compared to recent lattice results
for two colours \cite{Langfeld:2002}.
}
\end{figure}

The corresponding infrared behaviour of the propagators is given by
$D_{\text{Gluon}}(k^2) \sim (k^2)^{2 \kappa -1}$
and
$D_{\text{Ghost}}(k^2) \sim (k^2)^{-\kappa -1}$ with
$ \kappa  = (93 - \sqrt{1201})/98 \approx 0.595$:
One obtains a weakly infrared vanishing gluon propagator and a strongly
infrared enhanced ghost propagator. In Fig.\ \ref{lattice.dat} a comparison
of the numerical solution of DSEs with the results of recent lattice
calculations \cite{Langfeld:2002} for the case of two colours is shown.
As the solutions on the lattice include all
non-perturbative effects,  the results shown in Fig.~\ref{lattice.dat} suggest
that the omission of the two-loop diagrams in the truncated DSEs
mostly effects the region around the bending point at $1$ GeV. 
Given the limitations of both methods the qualitative and partly even 
quantitative agreement is remarkable. The combined evidence of the two 
methods points
strongly towards an infrared vanishing or finite 
gluon propagator and an infrared
singular ghost propagator in Landau gauge.

\section{Infrared fixed point and $\alpha_S(\mu^2)$}
\label{sec:IRfix}

The non-perturbative definition of the running coupling given in section 2 
can be summarized as follows:
\begin{equation}
\alpha_S(k^2) =  \alpha_S(\mu^2) Z(k^2;\mu^2)G^2(k^2;\mu^2).
\end{equation}
Here we have made  explicit the dependence of the propagator functions
on the renormalisation point.  An important point to notice in the results
described in the last section is the unique relation between the gluon and
ghost infrared behaviour. This is no accident: Consistency of the DSEs require
that the product $Z(k^2)G^2(k^2)$ goes to a constant in the infrared. 

Correspondingly we find an infrared fixed point of the running coupling:
\begin{equation}
\alpha_S(0)=\frac{2\pi}{3 N_c}
\frac{\Gamma(3-2\kappa)\Gamma(3+\kappa)\Gamma(1+\kappa)}{\Gamma^2(2-\kappa)
\Gamma(2\kappa)} \; , \quad \kappa =  \frac{93 - \sqrt{1201}}{98} \; .
\end{equation}
For the gauge group SU(3) the corresponding numerical value is 
$\alpha_S(0) \approx 2.972$. Of course, this result depends on the employed 
truncation scheme. In ref.\ \cite{Lerche:2001}, assuming the infrared dominance
of ghosts it has been shown that the tree-level vertex result
$\alpha_S(0) \approx 2.972$, among the general class 
of dressed ghost-gluon vertices considered in the infrared,
provides the maximal value for $\alpha_S(0)$.
If the exponent $\kappa$ is chosen
in an interval between 0.5 and 0.7 (as strongly suggested by lattice results)
one obtains $\alpha_S(0)> 2.5$ \cite{Lerche:2001}.

\begin{figure}
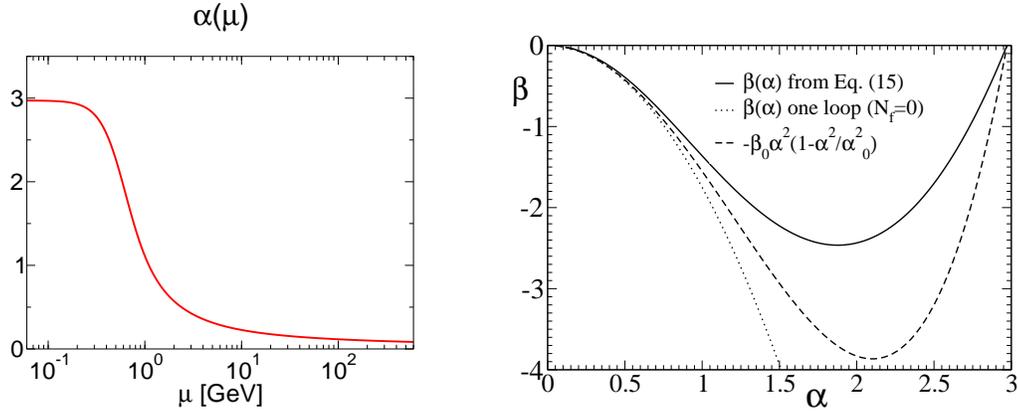

\centerline{
\epsfig{file=a_fit.eps,width=0.4\linewidth,height=0.4\linewidth}
\hfill
\epsfig{file=beta.eps,width=0.5\linewidth}
}
\caption{The strong running coupling (left panel) and the corresponding 
$\beta$ function (right panel, fully drawn line). The latter
is compared to the one-loop $\beta$ function (dotted line) and a 
polynomial in $\alpha$ (dashed line).   
\label{AlphaBeta}}
\end{figure}

In the case of three colours values for physical scales can be obtained  by 
requiring the experimental value $\alpha_S(M_Z^2=(91.2 \mbox{GeV})^2)=0.118$.
Together with the numerical solutions for the gluon and the ghost propagators
we can summarize our knowledge of the running strong coupling in the following 
fit:
\begin{eqnarray}
\alpha_S(x) &=& \frac{\alpha_S(0)}{\ln(e+a_1 x^{a_2}+b_1x^{b_2})},  \\
\alpha_S(0) &=& 2.972 , \quad 
a_1=5.292 \mbox{GeV}^{-2a_2}, \quad a_2=2.324 , 
b_1=0.034 \mbox{GeV}^{-2b_2}, \quad b_2=3.169 . \nonumber 
\end{eqnarray}
In fig.\ \ref{AlphaBeta} the running coupling and the corresponding $\beta$
function are shown.

Finally, we note that indications for an infrared finite coupling have
recently also been obtained within the background field method from the
Exact Renormalisation Group Equations \cite{Gies:2002af}.

\section{Outlook: $\alpha_S(\mu^2)$ and physical observables}
\label{sec:out}

To summarize: We have provided a non-perturbative definition for the running
strong coupling in the Landau gauge. The underlying picture is related to
confinement of transverse gluons, see {\it e.g.} \cite{vonSmekal:2000pz} 
and references therein. Loosely speaking, one can summarize this by stating
that gluons are confined by the Faddeev--Popov ghosts. The most interesting
result for the running coupling is the existence of an infrared fixed point.
The occurence of this fixed point can hereby be traced back to very general
properties of the ghost Dyson--Schwinger equation \cite{Lerche:2001}.
Taking only one-loop terms in the gluon Dyson--Schwinger equation into account
one obtains that $\alpha_S(\mu^2) \propto 1/N_c$ with $ N_c$ being the number
of colours. For three colours the corresponding critical value for
$\alpha_S(0)$ 
slightly depends on the approximations used. Together with the results of
lattice calculations we confidently conclude that $\alpha_S(0) \approx 3$
or slightly lower.

Of course, a phenomenological verification of the presented picture would be
highly welcome. From so-called analytic pertubation theory, see {\it e.g.}
\cite{Shirkov:2001sm} and references therein, it is known that an infrared
finite coupling allows for an almost straightforward calculation of amplitudes
relevant in inclusive $\tau$ decay, in $e^+$--$e^-$--annihilation into hadrons, 
in inelastic lepton-hadron scattering etc. The interesting point hereby is that
related observables might give bounds on allowed values for $\alpha_S(0)$.

\begin{ack}
R.A.~thanks the organizers of RENORMALIZATION GROUP 2002 for the possibility 
to participate in this extraordinarily interesting conference. 
The authors are grateful to J.~Bloch, H.~Reinhardt, D.~Shirkov, S.~Schmidt,
P.~Watson and D.~Zwanziger for helpful discussions. We are indebted to 
K.~Langfeld for communicating and elucidating his lattice results, 
partly prior to publication.

This work has been supported by the DFG under contract Al 279/3-3 and by the
European graduate school T\"ubingen--Basel (DFG contract GRK 683).
\end{ack}


\end{document}